\documentstyle[12pt]{article}
\renewcommand{\theequation}{\thesection.\arabic{equation}}
\newcounter{subequation}[equation]
\makeatletter
 
\expandafter\let\expandafter\reset@font\csname reset@font\endcsname
 
\def\subeqnarray{\arraycolsep1pt
    \def\@eqnnum\stepcounter##1{\stepcounter{subequation}%
        {\reset@font\rm(\theequation\alph{subequation})}}
\jot5mm     \eqnarray}

\makeatother\newcommand{\newsection}[1]{\vspace{10mm}
\pagebreak[3]\addtocounter{section}{1}\setcounter{equation}{0}
\setcounter{subsection}{0}\setcounter{footnote}{0}
 
\begin{flushleft}{\Large\bf \thesection. #1}
\end{flushleft}\nopagebreak\medskip\nopagebreak}

\def\beq{\begin{equation}}
\def\eeq{\end{equation}}
\def\eref#1{(\ref{#1})}
\newcommand\ftnote[1]{\setcounter{footnote}{#1}\addtocounter{footnote}{-1}
\footnote}
\def\tr{\mathop{\hbox{\rm tr}}\nolimits}
\def\e{{\rm e}}
\def\i{{\rm i}}
\def\d{{\rm d}}
\def\l{\lambda}
\def\s{\sigma}
\def\dd{\partial}
\def\half{\frac{1}{2}}
\def\?{(?)\marginpar{|?}}

\newfont{\bbd}{msbm10 scaled\magstep1}
\def\C{\bbd{C}}
\def\B{{\cal B}}
\def\F{{\cal F}}
\def\L{{\cal L}}
\def\M{{\cal M}}
\def\W{{\cal W}}
\def\SS{{\cal S}}
\def\JPA{J.\ Phys.\ A: Math.\ Gen.\ }
\def\IP{Inverse Problems}
\def\CMP{Commun.\ Math.\ Phys.\ }
\def\JMP{J.\ Math.\ Phys.\ }
\def\LMP{Lett.\ Math.\ Phys.\ }
\def\PLA{Phys.\ Lett.\ A }
\begin{document}
\begin{flushright}
\sf solv-int/9711010
\end{flushright}
\vskip1cm
\begin{center}\LARGE\bf
Few remarks on 
B\"{a}cklund transformations \\
for many-body systems
\end{center}
\vskip1cm
\begin{center}
V B Kuznetsov\ftnote{1}{On leave from: Department of Mathematical and
Computational Physics, Institute of Physics, St.~Petersburg University, 
St.~Petersburg 198904, Russia.
E-mail: {\tt vadim@amsta.leeds.ac.uk}}
and E K Sklyanin\ftnote{2}{On leave from: Steklov Mathematical Institute at 
St.~Petersburg, Fontanka 27, St.~Petersburg 191011, Russia. 
E-mail: {\tt sklyanin@pdmi.ras.ru}}
\vskip0.4cm
Department of Applied Mathematics,\\  
University of Leeds, Leeds LS2 9JT, UK
\end{center}
\vskip2cm
\begin{center}
\bf Abstract
\end{center}
 Using the $n$-particle periodic Toda lattice and the relativistic 
generalization due to Ruijsenaars of the elliptic
Calogero-Moser system as examples, we revise the basic properties of
the B\"acklund transformations (BT's) from the Hamiltonian point of view.
The analogy between BT and Baxter's quantum $Q$-operator pointed out by
Pasquier and Gaudin is exploited to produce a conjugated variable $\mu$
for the parameter $\l$ of the BT $B_\l$
such that $\mu$ belongs to the spectrum of the Lax operator $L(\l)$.
As a consequence, the generating function of the composition
$B_{\l_1}\circ\ldots\circ B_{\l_n}$ of $n$ BT's
gives rise also to another canonical transformation
separating variables for the model.
For the Toda lattice the dual BT parametrized by $\mu$ is introduced.
\vskip1cm
\noindent 14 November, 1997
\vskip1cm
\noindent Submitted to {\it\JPA}
\newpage

\newsection{Introduction}
\setcounter{equation}{0}
B\"{a}cklund transformations (BT's) are an important
tool in the theory of integrable systems \cite{BTbks}.
Most frequently, they are understood as special mappings between solutions
of nonlinear evolution equations. The Hamiltonian properties of BT's,
as canonical transformations, are studied less well. The recent developments
in the quantum integrable theories \cite{PG92,Qop},
discrete-time dynamics \cite{Ves91,NRK,FadVol} and separation of
variables \cite{SoV,KNS}
suggest, however, that the Hamiltonian aspect of BT's deserves more
attention.

The aim of the present
paper is to revise the concept of BT's from the Hamiltonian point of view and
to point out some new properties of BT's.
We restrict our attention to the finite-dimensional integrable
systems and illustrate our general remarks on the example of
the periodic Toda lattice and the elliptic Ruijsenaars model. 
When elaborating our approach to BT's, we have benefited greatly from the
works of Pasquier and Gaudin \cite{PG92}, where a fundamental relationship
between BT and Baxter's quantum $Q$-operator was discovered, and of Veselov 
\cite{Ves91},
who gave us the adequate mathematical language to speak about integrable
mappings.

In the section 2 the main properties of B\"{a}cklund transformations
for Liouville integrable systems are enlisted and a new property of
{\it spectrality} is introduced. The meaning of spectrality is elucidated
by making the comparison with the Baxter's quantum $Q$-operator. It is shown
that spectrality of BT provides an effective solution to the problem
of separation of variables. In two subsequent sections we illustrate
the new property of BT's for two families of integrable many-body 
systems. The concluding section 5 contains a summary and a discussion.

\newsection{Spectrality and separation of variables}
\setcounter{equation}{0}
Suppose an integrable system with $n$ degrees of freedom is described in
terms of the canonical Darboux variables $X\equiv\{X_i\}_{i=1}^n$ and
$x\equiv\{x_i\}_{i=1}^n$, with the Poisson brackets:
\beq
  \{X_i,X_j\}=\{x_i,x_j\}=0, \qquad \{X_i,x_j\}=\delta_{ij},
\eeq
and functionally independent commuting Hamiltonians $H_i\equiv H_i(X,x)$
\beq
     \{H_i,H_j\}=0, \qquad i,j=1,\ldots,n.
\eeq

For our purposes it is convenient to think of a BT as a canonical
transformation
$B_\l$ from the canonical variables $(X,x)$ to the canonical variables
$(Y,y)$. It is important that $B_\l$ depends on a complex parameter $\l$.
We shall suppose that $B_\l$ can be described via the generating function
$F_\l(y;x)$ such that
\beq
  X_i=\frac{\dd F_\l}{\dd x_i}, \qquad
  Y_i=-\frac{\dd F_\l}{\dd y_i}.
\label{eq:XYxy}
\eeq

The list of properties defining a BT usually includes:
\begin{itemize}
\item {\it Canonicity.} See above.
\item {\it Invariance of Hamiltonians.}
\beq
    H_i(X,x)=H_i(Y,y), \qquad i=1,\ldots,n.
\label{eq:inv-H-cl}
\eeq

\item {\it Commutativity.}
\beq
     B_{\l_1}\circ B_{\l_2}=B_{\l_2}\circ B_{\l_1}
\label{eq:com-b-cl}
\eeq
where $\circ$ means composition of canonical transformations.
\end{itemize}

In case of the {\it algebraically integrable} systems \cite{AvM89}
one more property can be added to the list:
\begin{itemize}
\item {\it Algebraicity.} The equations \eref{eq:XYxy} describing
$B_\l$ are supposed to be
algebraic with respect to $X$, $Y$ and properly chosen functions of
$x$ and $y$ (say, exponential or elliptic).
\end{itemize}

In the present paper, however, we concentrate on the analytic properties
of BT's and ignore their algebraic and algebro-geometric aspects.

It is important to make clear distinction between the notion of
BT and the close notions of {\it integrable canonical
mapping} \cite{Ves91}, or {\it integrable discrete-time dynamics}. The latter
two are defined by the properties of canonicity and 
invariance only, the parameter $\l$ being disregarded.
The term
`discrete-time dynamics' refers usually to the case when the canonical
transformation degenerates, in a certain limit, into an infinitesimal
generator $\{H,\cdot\}$ of a continuous Hamiltonian flow.
Existence of the parameter $\l$ is crucial for our definition
of BT and enriches it with new properties.

Though the commutativity of BT's is traditionally proved
as an independent property, in fact it follows from the canonicity and
the invariance of Hamiltonians. Indeed, as shown in \cite{Ves91},
any integrable canonical mapping acts on the Liouville torus as a shift
(or a collection of shifts, in case of multivalued mappings)
of the angle variables $\varphi_i\rightarrow\varphi_i+b_i(\l)$.
The commutativity is then obvious.

The theory of BT's acquires a new aspect if the integrable
system in question is solvable via Inverse Scattering (or Inverse
Spectral Transform) method.
Suppose that the commuting Hamiltonians $H_i$
can be obtained as the coefficients of the characteristic polynomial
\beq
   W(u,v;\{H_i\})=\det(v-L(u))
\eeq
of a matrix $L(u)\equiv L(u;X,x)$ (Lax operator) depending on $X$, $x$ and
a complex parameter $u$. Note that the invariance of $H_i$ under $B_\l$ is
equivalent then
to the invariance of the spectrum of $L(u)$, that is there
exists an invertible matrix $M(u)$ such that
\beq
   M(u)L(u;X,x)=L(u;Y,y)M(u),\qquad \forall u\in\C.
\eeq

The properties of BT's enlisted above are well known. Now we are
going to add to the list a new property which is the main contribution of
the present paper.
\begin{itemize}
\item {\it Spectrality.}
Let $\mu$ be defined as the variable conjugated to $\l$:
\beq
      \mu=-\frac{\dd F_\l}{\dd\l}.
\label{eq:def-mu}
\eeq

We shall say that the BT $B_\l$ is {\it associated} to the
Lax operator $L(u)$ if for some function $f(\mu)$ the pair $(\l,f(\mu))$
lies on the  {\it spectral curve} of the Lax matrix
\beq
      W(\l,f(\mu);\{H_i\})\equiv\det(f(\mu)-L(\l))=0.
\label{eq:spectr}
\eeq
\end{itemize}

This {\it spectrality} property of BT seems to be new, at least we failed
to find it in the literature. We have verified it for the Toda lattice and
the elliptic Ruijsenaars model for which $f(\mu)=\e^{-\mu}$ (see sections
3 and 4). It seems plausible, however, that spectrality is the property shared
by BT's for a much larger class of models.

The meaning of the equality \eref{eq:spectr} becomes clear if we
turn to the quantum case. In the pioneering paper by Pasquier and Gaudin 
\cite{PG92}, based on the earlier treatment of the classical Toda lattice
by Gaudin \cite{Gau83}, a remarkable connection has been established between
the classical BT $B_\l$ for the Toda lattice and the famous Baxter's 
$Q$-operator \cite{Bax82}. Pasquier and Gaudin have constructed certain
integral operator $\hat Q_\l$
\beq
    \hat Q_\l:\Psi(x)\rightarrow \int \d x\, Q_\l(y;x)\Psi(x)
\eeq
(here and below $\d x\equiv\d x_1\wedge\ldots\wedge\d x_n$ etc)
whose properties parallel those of the classical BT $B_\l$. In the quantum
case the canonical transformation is replaced with the similarity
transformation
\beq
   \hat Y_i=\hat Q_\l \hat X_i \hat Q_\l^{-1}, \qquad
   \hat y_i=\hat Q_\l \hat x_i \hat Q_\l^{-1},
\eeq
where the hat $\hat{\phantom{x}}$ distinguishes the quantum operators 
from their classical counterparts. 
The correspondence between the kernel $Q_\l(y;x)$ of $\hat Q_\l$ and the 
generating function $F_\l(y;x)$ of $B_\l$ is given by the semiclassical
relation
\beq
     Q_\l(y;x)\sim \exp\left(-\frac{\i}{\hbar}F_\l(y;x)\right), 
     \qquad \hbar\rightarrow0.
\eeq

After publication of \cite{PG92} the $Q$-operators have been found
for a number of other quantum integrable models \cite{Qop}.

The properties of $\hat Q_\l$ such as the invariance of the Hamiltonians
\beq
    [\hat Q_\l,H_i]=0
\eeq
and the commutativity
\beq
    [\hat Q_{\l_1},\hat Q_{\l_2}]=0
\eeq
reproduce the respective properties \eref{eq:inv-H-cl} and \eref{eq:com-b-cl}
of $B_\l$.
The most interesting property of $\hat Q_\l$, however, is that its eigenvalues
$\phi(\l)$ on the joint  eigenvectors $\Psi_\nu$ of $H_i$ and $Q_\l$
labelled with the quantum numbers $\nu$
\beq
    Q_\l\Psi_\nu=\phi_\nu(\l)\Psi_\nu
\label{eq:QPsi}
\eeq
satisfy the {\it separation equation}, which is a certain differential or
difference equation
\beq
    \hat W\left(\l,-\i\hbar\frac{\d}{\d\l};\{h_i\}\right)\phi_\nu(\l)=0
\label{eq:sepeq}
\eeq
containing the eigenvalues $h_i$ of $H_i$. In the classical limit the equation
\eref{eq:sepeq} goes over into the spectrality
equation \eref{eq:spectr}.

An important application of the spectrality property of BT is that to the
problem of {\it separation of variables} \cite{SoV,KNS}. Again, it is
instructive to start with the quantum case. A separating operator $\hat K$
is, by definition, an operator, transforming the joint eigenfunctions
$\Psi_\nu$ of $H_i$ into the product
\beq
    \hat K\Psi_\nu=c_\nu\prod_{i=1}^n \phi_\nu(\l_i)
\label{eq:K-sep-q}
\eeq
of {\it separated functions} $\phi_\nu(\l)$ of one variable $\l$
satisfying the separation equation \eref{eq:sepeq}. Since
the coefficients $c_\nu$ in \eref{eq:K-sep-q} can be chosen
arbitrarily, abstractly speaking,
there exist infinitely many separating operators $\hat K$.
The difficult problem, however, is to find the ones which
can be described as integral operators with explicitely given kernels.

Knowing a $Q$-operator gives one an immediate opportunity to construct plenty
of separating operators. Indeed, consider the operator product
$\hat Q_{\l_1\ldots\l_n}\equiv\hat Q_{\l_1}\ldots\hat Q_{\l_n}$
having the kernel $Q_{\l_1\ldots\l_n}(y;x)$ and for any function
$\rho(y)$ introduce the operator
\beq
   \hat K_\rho:\Psi(x)\rightarrow\int \d x \int \d y\,
   \rho(y)Q_{\l_1\ldots\l_n}(y;x)\Psi(x).
\label{eq:def-K}
\eeq

It is obvious from \eref{eq:QPsi} that $\hat K_\rho$ is a separating operator,
the coefficients $c_\nu$ being
\beq
    c_\nu=\int\d y\,\rho(y)\Psi_\nu(y).
\label{eq:c-rho}
\eeq

Since the eigenfunctions $\Psi_\nu(y)$ form a basis in the corresponding
Hilbert space, the formula
\eref{eq:c-rho} provides a one-to-one correspondence between reasonably chosen
classes of $c_\nu$ and $\rho(y)$. Therefore, arguably,
the formula \eref{eq:def-K} describes all possible separating operators.
Their kernels $K_\rho(\l;x)$ are given explicitely as multiple integrals
\begin{eqnarray}
K_\rho(\l;x)&=&\int\d{y}\int\d\xi^{(1)}\ldots\int\d\xi^{(n-1)} 
\nonumber \\
&&  \times\rho(y)Q_{\l_1}(y;\xi^{(1)})Q_{\l_2}(\xi^{(1)};\xi^{(2)})\ldots
  Q_{\l_n}(\xi^{(n-1)};x).
\label{eq:def-K-ker}
\end{eqnarray}

It is a straightforward task to present the classical analog of the above
argument. Consider the composition
$B_{\l_1\ldots\l_n}=B_{\l_1}\circ\ldots\circ B_{\l_n}$ of B\"{a}cklund
transformations and the corresponding generating function
$F_{\l_1\ldots\l_n}(y;x)$. Let us switch now the roles of $y$'s and $\l$'s
treating $\l$'s as dynamical variables and $y$'s as parameters. Then
$F_{\l_1\ldots\l_n}(y;x)$  becomes the generating function of the
$n$-parametric
canonical transformation $K_y$ from $(X,x)$ to $(\mu,\l)$ given by
the equations
\beq
  X_i=\frac{\dd F_{\l_1\ldots\l_n}}{\dd x_i}, 
\qquad \mu_i=-\frac{\dd F_{\l_1\ldots\l_n}}{\dd\l_i}.
\label{eq:K-canon}
\eeq

It follows directly from \eref{eq:spectr} that the pairs $(\l_i,\mu_i)$
satisfy the separation equations
\beq
     W(\l_i,f(\mu_i);\{H_j\})=0
\eeq
which constitutes exactly the definition of the separating canonical
transformation in the classical case \cite{SoV}.

The above construction corresponds in the quantum case to setting
$\rho(y)=\delta(y_1-\bar{y}_1)\ldots\delta(y_n-\bar{y}_n)$
where $\bar{y}_i$ are some constants.
It remains an open question what could be the
classical analog of the formula \eref{eq:def-K} for generic $\rho(y)$.

As the last general remark before passing to the examples, we would like to
stress that for the finite-dimensional systems the composition of $n$
BT's with $n$ being the number of degrees of freedom is a sort of
`universal' BT in the sense that any other canonical transformation
preserving the Hamiltonians $H_i$ must be expressible in terms of
$B_{\l_1\ldots\l_n}$. To observe it one can use again the fact that
in the angle coordinates $B_\l$ acts as a shift
$\varphi_i\rightarrow\varphi_i+b_i(\l)$. For generic $b_i(\l)$ the
sum $b_i(\l_1)+\ldots+b_i(\l_n)$ must then cover the
$n$-dimensional Liouville torus which results in the
universality of $B_{\l_1\ldots\l_n}$.

\newsection{Periodic Toda lattice}
\setcounter{equation}{0}
Our first example is the periodic Toda lattice \cite{Toda,Moe76} 
for which there exist two
alternative Lax operators associated, as we shall show, with two different
BT's. The standard and quite well studied BT \cite{HS,Toda,BTbks,Gau83}
which we
denote here $B_\l$ is associated, in the sense defined in the previous
section, to the $2\times2$ Lax matrix (or, monodromy matrix \cite{FTbook}) 
$L(u;X,x)$ defined as the product of local $L$-operators
\beq
 L(u)=\ell_n(u)\ldots\ell_2(u)\ell_1(u),
\label{eq:toda-Lll}
\eeq
\beq
 \ell_i(u)\equiv\ell_i(u;X_i,x_i)=\left(\begin{array}{cc}
     u+X_i & -\e^{x_i} \\
     \e^{-x_i} & 0 \end{array}\right).
\label{eq:def-ell}
\eeq

The characteristic polynomial of $L(u)$ is quadratic in $v$
\beq
  W(u,v)\equiv\det(v-L(u))=v^2-t(u)v+1,
\eeq
and the commuting Hamiltonians $H_i$ are obtained from the expansion of the 
only non-trivial spectral invariant $t(u)\equiv \tr L(u)$
\beq
 t(u)=u^n+H_1u^{n-1}+\ldots+H_n.
\eeq

In particular,
\beq
 \half H_1^2-H_2=\sum_{i=1}^n \left(\half X_i^2+\e^{x_{i+1}-x_i}\right)
\eeq
is the standard periodic Toda Hamiltonian (in this section we use the 
periodicity convention $i+n\equiv i$ for the indices $i$).

The B\"acklund transformation $B_\l$ is obtained from the generating function
\beq
 F_\l(y;x)=\sum_{i=1}^n \bigl(\e^{x_i-y_i}-\e^{y_{i+1}-x_i}-\l(x_i-y_i)\bigr)
\label{eq:F-toda}
\eeq
and, according to \eref{eq:XYxy}, is implicitely described by the equations
\beq
 X_i=\e^{x_i-y_i}+\e^{y_{i+1}-x_i}-\l, \qquad
 Y_i=\e^{x_i-y_i}+\e^{y_i-x_{i-1}}-\l.
\label{eq:XY-toda}
\eeq

The characteristic properties of the BT are verified easily.
The {\it invariance of the Hamiltonians} can be established using
the equality \cite{Gau83}
\beq
 M_{i+1}(u,\l)\ell_i(u;X_i,x_i)=\ell_i(u;Y_i,y_i)M_i(u,\l),
\label{eq:gauge-toda}
\eeq
where
\beq
 M_i(u,\l)\equiv M_i(u,\l;x_{i-1},y_i)
 =\left(\begin{array}{cc}
    1 & -\e^{y_i} \\
   \e^{-x_{i-1}} & \l-u-\e^{y_i-x_{i-1}} \end{array}\right)
\label{eq:def-M-toda}
\eeq
which one can verify directly using the equations
\eref{eq:XY-toda}. Due to the periodic boundary conditions, the local gauge
transformation \eref{eq:gauge-toda} results in the spectrum-preserving
similarity transformation
\beq
  M_1(u,\l)L(u;X,x)=L(u;Y,y)M_1(u,\l)
\label{eq:M1L-toda}
\eeq
for $L(u)$ which proves the invariance \eref{eq:inv-H-cl} of the Hamiltonians.

The direct proof of the {\it commutativity} \eref{eq:com-b-cl} of the BT's
can be found in \cite{HS,Toda,BTbks}.

To prove the {\it spectrality} equality \eref{eq:spectr}
which in this case takes the form $\det(\e^{-\mu}-L(\l))=0$
we shall apply a modified version of the argument used in \cite{Gau83,PG92}
for the quantum case.
Note, first, that in our case
\beq
 \mu=-\frac{\dd F_\l}{\dd\l}=\sum_{i=1}^n (x_i-y_i),
\label{eq:def-mu-toda}
\eeq
as follows from \eref{eq:F-toda} and \eref{eq:def-mu}. It suffices then to
show that $\e^{-\mu}$ is an eigenvalue of the matrix $L(\l)$. We shall
construct explicitely the corresponding eigenvector $\omega_1$:
\beq
 L(\l;X,x)\omega_1=\e^{-\mu}\omega_1.
 \label{eq:Lom-toda}
\eeq

{}From \eref{eq:def-M-toda} it follows that
$\det\bigl(M_i(u,\l)\bigr)=\l-u$. It is easy to see that
for $u=\l$ the matrix $M_i(\l,\l)$ has the unique, up to a scalar
factor, null-vector
\beq
\omega_i=\left(\begin{array}{c}\e^{y_{i}} \\ 1 \end{array}\right), \qquad
  M_i(\l,\l)\omega_i=0.
\eeq
Using the identity \eref{eq:M1L-toda} we conclude that
\beq
M_1(\l,\l)L(\l;X,x)\omega_1=0
\eeq
which, combined with the uniqueness of the null-vector $\omega_1$ of
$M_1$, implies that $\omega_1$ is an eigenvector of $L(\l;X,x)$. To
determine the corresponding eigenvalue, we apply the same argument
to the identity \eref{eq:gauge-toda} obtaining the equality
$M_{i+1}(\l;\l)\ell_i(\l;X_i,x_i)\linebreak[0]\omega_i=0$ 
from which it follows that
$\ell_i(\l;X_i,x_i)\omega_i\sim\omega_{i+1}$. The direct calculation shows
that
\beq
 \ell_i(\l;X_i,x_i)\omega_i=\e^{y_i-x_i}\omega_{i+1}.
\label{eq:vacuum-ell}
\eeq
It remains only to use the formulae \eref{eq:toda-Lll} and
\eref{eq:def-mu-toda} to arrive finally at \eref{eq:Lom-toda}.
Actually, we could skip the discussion of null-vectors of $M_i$ and
to derive \eref{eq:Lom-toda} directly from \eref{eq:vacuum-ell}.
In more complicated situations, however, it may be easier to find $\omega$
as the null-vector of $M$ and then to determine the corresponding eigenvalue
of $L(\l)$.

Note that the vectors $\omega_i$ are the classical counterparts of
Baxter's \cite{Bax82} vacuum vectors.

Let us examine now the alternative Lax operator \cite{Toda,Moe76}
given by the $n\times n$ matrix $\L(v;X,x)$ with the components
\beq
 \L_{jk}(v;X,x)=
-X_j\delta_{jk}+v^{-1/n}\e^{x_j-x_k}\delta_{j,k+1}+v^{1/n}\delta_{j+1,k}.
\label{eq:dual-L-toda}
\eeq

The duality between the Lax operators $\L(v)$ and $L(u)$ is expressed in the
switching the roles of the parameters $u$ and $v$.
The characteristic polynomial $\W(v,u)\equiv\det(u-\L(v))$ of the Lax operator
\eref{eq:dual-L-toda} produces the same Hamiltonians $H_i$
and the same spectral curve as $W(u,v)$, as follows from the identity
\beq
  \det(v-L(u))=-v\det(u-\L(v)).
\eeq

For other examples of the similar duality, see \cite{dual}.

The swapping of $u$ and $v$ corresponds to switching the roles of the
parameters $\l$ and $\mu$ in the BT. For the new B\"acklund transformation
$\B_\mu$ associated with the Lax operator
$\L(v)$ the formulae \eref{eq:F-toda}, \eref{eq:XY-toda} and
\eref{eq:def-mu-toda} remain the same but their interpretation changes.
The BT is parametrized now by the parameter $\mu$ which becomes a numerical
constant. The equality \eref{eq:def-mu-toda}  is reinterpreted now as a
constraint on the variables
$x_i$ and $y_i$. The parameter $\l$ is reinterpreted,
respectively, as the Lagrange multiplier for the constraint 
\eref{eq:def-mu-toda} and
becomes a dynamical variable which can be defined from the equations
\eref{eq:XY-toda}.

The characteristic properties of BT are verified for $\B_\mu$ in very
much the same manner like for $B_\l$.
The {\it invariance of the Hamiltonians} follows from the
invariance of the spectrum of $\L(v)$ which, in turn, follows from the
easily verified identity
\beq
   \M(v)\L(v;X,x)=\L(v;Y,y)\M(v),
\eeq
with the matrix $\M(v)\equiv\M(v;x,y)$ given by its components
\beq
 \M_{jk}(v)=-\delta_{jk}+v^{-1/n}\e^{y_j-x_k}\delta_{j,k+1}.
\eeq

The {\it commutativity}, as shown in section 2 follows from the canonicity
and the invariance.

To prove the {\it spectrality} equality  $\det(\l-\L(\e^{-\mu}))=0$,
it suffices, similarly to the case of the Lax operator $L(u)$,
to present the eigenvector $\Omega$ of the matrix $\L(\e^{-\mu})$
corresponding to the eigenvalue $\l$:
\beq
 \L(\e^{-\mu})\Omega=\l\Omega.
\label{eq:eigv-toda}
\eeq

Again, $\Omega$ can be determined as the null-vector of $\M(\e^{-\mu})$:
\beq
 \M(\e^{-\mu})\Omega=0.
\label{eq:toda-MOm=0}
\eeq
Note that the uniqueness of $\Omega$ follows from the easily verified identity
$\det\bigl(z-\M(v)\bigr)=(z+1)^n-v^{-1}\e^{-\mu}$ which implies that the
spectrum of $\M(\e^{-\mu})$ consists of
$n$ non-degenerate eigenvalues, the 0 being
one of them. From \eref{eq:toda-MOm=0} one easily derives the
recurrence relation for the components of $\Omega$
\beq
 \Omega_j=\Omega_{j-1}\exp\left(y_j-x_{j-1}+\frac{\mu}{n}\right)
\eeq
which determine $\Omega$ up to a constant factor. It remains to
verify the identity \eref{eq:eigv-toda} which can be done by
a direct calculation using
the expressions \eref{eq:dual-L-toda} for the matrix $\L(v)$,
\eref{eq:def-mu-toda} for $\mu$ and \eref{eq:XY-toda} for $X_i$.

\newsection{Elliptic Ruijsenaars model}
\setcounter{equation}{0}
Our second example is the relativistic generalization due to Ruijsenaars
\cite{Ruij87} of the elliptic Calogero-Moser \cite{Cal} many-body system.
For the non-relativistic Calogero-Moser system a BT was found in
\cite{CMbckl}. In \cite{NRK}
a discrete-time dynamics was constructed for the elliptic Ruijsenaars model.
As we show below, the discrete-time evolution transformation found in
\cite{NRK} has all the properties of a BT if the parameter $p$ in \cite{NRK} 
is specified in a proper way.

We use here the notations of \cite{NRK} with few exceptions: our parameter
$\xi$ equals to $-\l$ from \cite{NRK}, and our $\e^{-\l}$ corresponds to $p$
from \cite{NRK}. As in the case of the Toda lattice, there exist two
dual BT's: $B_\l$ and $\B_\mu$. The standard Lax operator for the Ruijsenaars
model, as shown below, is associated with $\B_\mu$. Since the dual Lax
operator is so far unknown, we describe here only the transformation $\B_\mu$.

Following \cite{NRK,KNS}, we introduce the Lax operator $\L(v;X,x)$ 
for the $n$-particle ($A_{n-1}$ type) Ruijsenaars system as
the $n\times n$ matrix with the entries
\beq
 \L_{ij}(v)=-\e^{X_i}\frac{\s(\xi)\s(v+x_i-x_j-\xi)}%
{\s(v)\s(x_i-x_j-\xi)}
 \prod_{k\neq i}\frac{\s(x_i-x_k+\xi)}{\s(x_i-x_k)},
\label{eq:Lij-Ruijs}
\eeq
where $\s(x)$ is the Weierstrass sigma function and $\xi$ is a constant.

The commuting Hamiltonians 
\beq
H_i=\sum_{J\subset\{1,\ldots,n\} \atop \left|J\right|=i}
\exp\left(\sum_{j\in J}X_j\right)\,
  \prod_{j\in J \atop k\in \{1,\ldots,n\}\setminus J} 
\frac{\s(x_j-x_k+\xi)}{\s(x_j-x_k)},
\qquad i=1,\ldots,n
\label{eq:def-H}\eeq
are generated from the characteristic polynomial of the matrix $\L(v)$ 
(\ref{eq:Lij-Ruijs})
\beq
\det(\L(v)-u)=\sum_{j=0}^n (-u)^{n-j}H_j\,
\frac{\s(v-j\xi)}{\s(v)}
\label{eq:char-poly-N}
\eeq
where we assume $H_0\equiv1$.

The B\"acklund transformation $\B_\mu$  is given by the equations
\beq
 \e^{X_i}=\e^{-\l}\prod_{j\neq i}\frac{\s(x_i-x_j-\xi)}{\s(x_i-x_j+\xi)}
 \prod_{k=1}^n\frac{\s(x_i-y_k+\xi)}{\s(x_i-y_k)}
\label{eq:Xi-Ruij}
\eeq
\beq
 \e^{Y_i}=\e^{-\l}\prod_{k=1}^n\frac{\s(x_k-y_i+\xi)}{\s(x_k-y_i)}
\label{eq:Xi-Ruij-x}
\eeq
where $\l$ is considered as the Lagrange multiplier corresponding to the
constraint
\beq
 \mu=n\xi+\sum_{k=1}^n(x_k-y_k).
\label{eq:def-mu-Ruij}
\eeq
Note here that the variable $\l$ in formulas 
(\ref{eq:Xi-Ruij})--(\ref{eq:Xi-Ruij-x}), describing the 
dicrete-time dynamics, appeared already as $p$ in 
\cite{NRK}, but the conjugated variable $\mu$ did not.
Notice also that $\l$ was treated in \cite{NRK} as an 
extra parameter, not as a Lagrange multiplier 
corresponding to a constraint.

The generating function of the canonical transformation $\B_\mu$
is expressed in terms of the function
\beq
  \SS(x)=\int^x\ln\s(y)dy
\eeq
as follows:
\begin{eqnarray}
\F_\l(y;x)&=&-\l\sum_{i=1}^n(x_i-y_i+\xi)
+\sum_{i<j}\bigl(\SS(x_i-x_j-\xi)-\SS(x_i-x_j+\xi)\bigr) \nonumber \\
&&  +\sum_{i,j=1}^n\bigl(\SS(x_i-y_j+\xi)-\SS(x_i-y_j)\bigr).
\end{eqnarray}

The verification of the characteristic properties of BT for $\B_\mu$
proceeds in the same way as in the case of the Toda lattice.

The {\it invariance of the Hamiltonians $H_i$} follows from the
identity (see \cite{NRK} for the proof)
\beq
   \M(v)\L(v;X,x)=\L(v;Y,y)\M(v)
\eeq
where the matrix $\M(v)\equiv\M(v;x,y)$ is defined as
\beq
  \M_{ij}(v)=
  \frac{\s(v+y_i-x_j-\xi)}{\s(y_i-x_j-\xi)}
  \prod_{k\neq i}\frac{\s(y_i-y_k+\xi)}{\s(y_i-y_k)}
  \prod_k\frac{\s(x_k-y_i+\xi)}{\s(x_k-y_i)}.
\eeq

The {\it commutativity}, as usual, is a consequence of canonicity
and invariance (see section 2).

To prove the {\it spectrality} equality which takes the form
$\det(\e^{-\l}-\L(\mu))=0$ it is sufficient, like in the case of the Toda
lattice, to find the eigenvector $\Omega$ of the matrix $\L(\mu)$
corresponding to the eigenvalue $\e^{-\l}$. Let us show that, up to a
constant multiplier, the components of the eigenvector $\Omega$ are
\beq
   \Omega_i=\frac{\prod\limits_{k=1}^n\s(x_i-y_k+\xi)}%
{\prod\limits_{k\neq i}\s(x_i-x_k)}.
\label{Omega}
\eeq

The equality
\beq
     \L(\mu)\Omega=\e^{-\l}\Omega
\eeq
or
\beq
     \sum_{j=1}^n \L_{ij}(\mu)\Omega_j=\e^{-\l}\Omega_i
\eeq
after the substitutions \eref{eq:Lij-Ruijs} for $\L_{ij}$,
\eref{eq:def-mu-Ruij} for $\mu$ and \eref{eq:Xi-Ruij} for $\e^{X_i}$
is reduced to the following identity for sigma functions
\beq
  \sum_{j=1}^n\s(\mu+x_i-x_j-\xi)\prod\limits_{k=1}^n\s(x_j-y_k+\xi)
\prod\limits_{k\neq j}\frac{\s(x_i-x_k-\xi)}{\s(x_j-x_k)}
=\s(\mu)\prod_{k=1}^n\s(x_i-y_k).
\label{eq:ident-s}
\eeq

Due to the symmetry, it is sufficient to prove \eref{eq:ident-s} only for
$i=1$. Let $i=1$ and $n\geq2$.
Consider the both sides of the equality \eref{eq:ident-s} as 
functions of $x_n$. It is easy to see that they are holomorphic in $x_n$
(the apparent poles in the left-hand-side being cancelled) and have the
same quasiperiodicity properties. From the basic properties of sigma
functions \cite{WW} it follows that it is sufficient to verify the equality
of LHS and RHS only in one arbitrary point $x_n=\bar x$
with the only condition $\bar x\neq\mu-x_n$. Choosing
$\bar x=y_n-\xi$ we observe that \eref{eq:ident-s} is reduced to the 
similar identity of order $n-1$. The proof follows then by induction in $n$
since the case $n=1$ is trivial.

As in the section 3, the vector $\Omega$ (\ref{Omega}) is again
the null-vector of the matrix $\M(\mu)$, i.e.
\beq
\M(\mu)\Omega=0.
\label{21}
\eeq

The corresponding identity for sigma functions
\beq
\sum_{j=1}^n\s(\mu+y_i-x_j-\xi)\;\frac{\prod_{k\neq i}\s(x_j-y_k+\xi)}
{\prod_{k\neq j}\s(x_j-x_k)}=0
\label{22}
\eeq
follows from the identity
\beq
\sum_{j=1}^n 
\frac{\textstyle \prod_{k=1}^n \sigma(x_j-z_k)}%
{\textstyle \prod_{k\neq j}\sigma(x_j-x_k)}\,=\,0\   \quad
{\rm if} \quad 
~~  \ \sum_{k=1}^n (z_k-x_k) = 0\,, 
\eeq
(cf. \cite{WW}, p. 451) when one substitutes $z_k=y_k-\xi$ for $k\neq i$
and $z_i=\mu+y_i-\xi$.

\newsection{Discussion}
\setcounter{equation}{0}
We have studied
three new aspects of B\"{a}cklund transformations. 
Those are spectrality, dual BT's and
application of BT's to the problem of separation of variables.
As demonstrated in the section 2, the composition of 
$n$ BT's, being an `universal' ($n$-parametric) BT,
provides a separation of variables which has $n$
arbitrary parameters, and thereby defines an `universal'
($n$-parametric) family of separating transformations.
The connection between the 
`universal' BT and the `universal' SoV is intriguing and has yet
to be studied in detail.

Though we have discussed in the present paper only the classical case,
our primary motivation comes from the quantum case. The main problem
in the quantum case is to construct Baxter's $Q$-operator which
is a quantum analog of the B\"{a}cklund transformation.
For the trigonometric case of the 
Ruijsenaars system, i.e. for the case of multivariable ($A_{n-1}$-type) 
Macdonald polynomials, we have succeeded to describe explicitely
such a quantum analog of the transformation
$\B_\mu$ introduced in the section 4. The results will be reported elsewhere.

\section*{Acknowledgements}

We are grateful to J Gibbons, I V Komarov and F W Nijhoff for
their interest in the work and useful discussions. The authors 
wish to acknowledge the support of EPSRC.



\begin{thebibliography}{12}

\bibitem{BTbks} Rogers C and Shadwick W F 1982 {\it B\"{a}cklund
Transformations and Their Applications} (New York: Academic Press) 
\item[]Matveev V B and Salle M A 1991 {\it Darboux
transformations and solitons} (Berlin: Springer)

\bibitem{PG92} Pasquier V and Gaudin M 1992 The periodic Toda chain
and a matrix generalization of the Bessel function recursion relations
{\it\JPA\ }{\bf 25} 5243--5252

\bibitem{Qop} Antonov A and Feigin B 1996 Quantum group representations and 
Baxter equation {\it Preprint} hep-th/9603105
\item[] Bazhanov V V, Lukyanov S L and Zamolodchikov A B 1996 
Integrable structure of conformal field theory 
II. $Q$-operator and DDV equation {\it Preprint} hep-th/9604044

\bibitem{Ves91} Veselov A P 1991 Integrable maps {\it
Russian Math.\ Surveys }{\bf 46}:5  1--51

\bibitem{NRK} Nijhoff F W, Ragnisco O and Kuznetsov V B 1996
Integrable time-discretization of the Ruijsenaars-Schneider model
{\it\CMP }{\bf 176} 681--700

\bibitem{FadVol} Faddeev L D and Volkov A Yu 1994 Hirota equation as an
example of integrable symplectic map {\it\LMP} {\bf 32} 125--136

\bibitem{SoV} Sklyanin E K 1995 Separation of variables. New trends.
{\it Progr.\ Theor.\ Phys.\ Suppl.\ }{\bf 118} 35--60  
\item[] Kuznetsov V B and Sklyanin E K 1996 Separation of variables 
for the $A_2$ Ruijsenaars model and a new integral representation for
the $A_2$ Macdonald polynomials {\it\JPA\ }{\bf 29} 2779--2804

\bibitem{KNS} Kuznetsov V B, Nijhoff F W and Sklyanin E K 1997
Separation of variables for the Ruijsenaars system {\it Preprint}
solv-int/9701004; to appear in {\it \CMP}

\bibitem{AvM89} Adler M and van Moerbeke P 1989
{\it Algebraic integrable systems: A systematic approach}
(Boston: Academic Press)

\bibitem{Gau83} Gaudin M 1983 {\it La Fonction d'Onde de Bethe}
(Paris: Masson)

\bibitem{Bax82} Baxter R I 1982 {\it Exactly Solved Models in
Statistical Mechanics} (London: Academic Press, London) chapters 9--10

\bibitem{Toda} Toda M 1981 {\it Theory of Nonlinear Lattices}
(Berlin: Springer-Verlag)

\bibitem{Moe76} van Moerbeke P 1976 The spectrum  of Jacobi matrices
{\it Invent.\ Math.\ } {\bf 37} 45--81

\bibitem{HS} Hirota R and Satsuma J 1978 A simple structure of
superposition formula of the B\"{a}cklund transformation
{\it J.\ Phys.\ Soc.\ Japan }{\bf 45} 1741--1750

\bibitem{FTbook} Faddeev L D and Takhtajan L A 1987
{\it Hamiltonian Methods in the Theory of solitons} (Berlin: Springer-Verlag)

\bibitem{dual} Adams M R, Harnad J and Hurtubise J 1990 Dual moment maps to
loop algebras {\it\LMP} {\bf 20} 294--308
\item[] Nijhoff F W, Papageorgiou V G, Capel H W and Quispel G R W
1992 The lattice Gel'fand-Dikii hierarchy {\it\IP\ }{\bf 8} 597--621

\bibitem{Ruij87} Ruijsenaars S N M 1987 Complete integrability of
relativistic Calogero-Moser systems and elliptic function identities
{\it\CMP\ }{\bf 110} 191--213

\bibitem{Cal} 
Calogero F 1971 Solution of the one-dimensional $N$-body problems 
with quadratic and/or inversely quadratic pair potentials
{\it \JMP\ }{\bf 12} 419--436
\item[]
Moser J 1975 Three integrable Hamiltonian systems connected with 
isospectral deformations {\it Adv. Math.\ }{\bf{16}} 197--220
\item[]
Olshanetsky M A and Perelomov A M 1981 Classical integrable 
finite-dimensional systems related to Lie 
algebras {\it Phys. Rep.\ }{\bf 71} 313--400

\bibitem{CMbckl} Chudnovsky D V and Chudnovsky G V 1977 Pole expansions
of nonlinear partial differential equations {\it Nuovo Cimento B }{\bf 40}
339--353
\item[] Gibbons J 1982 Rational solutions of Lax equations
{\it Unpublished}
\item[] Wojciechowski S 1982 The analogue of the
B\"{a}cklund transformation for integrable many-body systems
{\it \JPA\ }{\bf 15} L653--657; 1983 Corrigendum {\bf 16} 671
\item[] Gibbons J, Hermsen T and Wojciechowski S 1983
A B\"{a}cklund transformation for a generalized Calogero-Moser system
{\it \PLA\ }{\bf 94} 251--253
\item[] Nijhoff F W and Pang G D 1996 Discrete-Time
Calogero-Moser Model and Lattice KP Equations {\it Proc.\ of the Intnl.\
Workshop on Symmetries and Integrability of Difference Equations}
eds D Levi {\it et al} (CRM Proceedings and Lecture Notes
{\bf 9}) 253--264

\bibitem{WW} Whittaker E T and Watson G N 1988 
{\it A course in modern analysis} (Cambridge University 
Press, 4th ed.)

\end{thebibliography}
\end{document}